# A correlation between drag and an integral property of the wake

T. S. MORTON

An integral quantity is presented that relates the wake of a body in nominally two-dimensional flow to its drag, for Reynolds numbers ranging from 9,000 to 144,000. It is defined as the ratio of the kinetic energy to the vorticity in the fluid boundary and, for the special case of laminar flow, is proportional to the angular momentum in the wake bubble. The new quantity is useful for correlating drag data for circular and rectangular cylinders, wedges, v-gutters, and normal flat plates with and without splitter plates. The correlation indicates that the drag force is proportional to the flow speed and the mass flow rate stored in the boundary of the fluid, where the fluid boundary is defined so as to include the wake bubble. Order-of-magnitude arguments indicate that, absent any quantization of vortex size, this mass flow rate, and hence the drag force, can become unbounded as the vortices contained in the wake becomes finer.

## 1. Introduction

This paper proposes a new quantity useful in correlating aerodynamic drag of bluff bodies to flow properties of the near wake in the range $(9 \times 10^3 \le Re \le 1.44 \times 10^5)$, originally reported by Morton (1997). In the context of bluff-body flows, the quantity is the ratio of the kinetic energy (mean, periodic, and turbulent) to the vorticity in the wake of the body.

No general theory is available that consistently matches drag data. The long distances traveled by golf balls have been generally explained in terms of high momentum transport from the turbulent boundary layer (see Davies 1949). The drag reducing capability of riblets (see Walsh and Lindemann 1984, Coustols 1996) has likewise been explained in similar terms. However, Djenidi *et al.* (1994) found that appropriately sized riblets reduce frictional drag in *laminar* flows as well, at least on a per unit of wetted length basis. The most generally accepted explanation is that conditions along the upstream boundary layer, such as surface roughness, increase turbulence and thereby reduce drag (see, e.g., Granger 1995, p. 785). Bloor (1964), however, found that as Reynolds number increases, turbulence creeps into the wake towards the body from far downstream rather than from upstream. Moreover, the measurements of Cantwell and Coles (1983) show that the entire wake can be made fully turbulent and yet the upstream boundary layer remain laminar and forward on the body.

Early discussions on the enhanced flying characteristics of spinning projectiles, enabled, for example, by the rifling of a gun barrel, focused primarily on gyroscopic stability (Milne-Thomson 1968 p. 558; Thwaites 1960, p. 418). Improvements to the aerodynamic characteristics of the spinning Frisbee have been attributed to spoilers that "disrupt the airflow" and "create a turbulent unseparated boundary layer over the convex side of the saucer and… result in a reduction of drag especially in high-speed flight and an increase in stability while in flight" (Headrick 1965). This description parallels the classical explanation of drag reduction wherein the drag coefficient of a body is reduced due to a tripping of the boundary layer to turbulent flow via high momentum mixing. This added momentum is said to cause the boundary layer to remain attached farther downstream on the body surface. It is clear that a lack of separation improves the drag characteristics of a body. The question remains as to what causes the lack of separation. The phrase "high momentum mixing," often used to explain this phenomenon, has not generally proved to be extremely useful from a predictive standpoint. More recent ring designs, based on the Frisbee concept, with a bulk of material along the outer radius, appear to have even smaller drag coefficient. The enhanced flying characteristics of these designs are often attributed to the "spoiler" material on the outer rim. However, an equal, if not greater, effect of the massive outer rim is likely the prolonged rotation sustained by its flywheel affect. As noted by Potts and Crowther (2002), the distance record for a thrown Frisbee now stands at 250 m. The distance record for any hand-thrown object was set at 406.29 m in 2003 using the Aerobie Pro Ring and then at 427.2 m by Schummy (2005) using a boomerang. It seems unlikely that any non-rotating object could be thrown such distances, regardless of its shape or stability. Compare the above facts, for example, with the distance record for the javelin set at 98.48 m in 1996.

One of the more subtle but curious contributions provided by CFD studies in the last few decades has been the placing on fairly firm footing the concept observable in the experimental data presented by Swanson (1961) and Tanaka and Nagano (1972), namely that under certain conditions, rotation of a cylinder not only generates lift but also reduces drag (see, e.g., Ingham 1983, Diaz *et al.* 1985, Badr *et al.* 1989, Ingham and Tang 1990, D'Alessio and Dennis 1994, Kang and Choi 1999, Stojković *et al.* 2002, Stojković *et al.* 2003, Inoue *et al.* 2003, Mittal and Kumar 2003).

The major difficulty in theoretically predicting the flow properties for bluff-body flow problems lies in the separated boundary layer. Specific problems encountered when trying to apply conventional boundary layer theory to separated boundary layers include: a lack of knowledge of the proper far stream pressure boundary condition along the region leading up to separation, uncertainty as to the location of the point of separation, and uncertainty as to the pressure at which separation occurs.

Models of the wake aimed at predicting drag are numerous (Kirchhoff 1869, Joukowsky 1890, Riabouchinsky 1920, McNown and Yih 1953, Roshko 1954, Lavrentiev 1962, Wu 1962, Parkinson and Jandali 1970, Wu 1972, Yeung and Parkinson 1997). These rely heavily on a knowledge of the shape of the body in question. The closed wake model suggested by Batchelor (1956) focused primarily on finding the external flow past regions of uniform vorticity (see Childress 1966, Sadovskii 1971, Turfus 1993). Roshko (1993) derived an expression for the length of the mean wake bubble in terms of the base suction coefficient and Reynolds shear stress for a normal flat plate with a trailing splitter plate. Using the free streamline model, Roshko was then able to predict a reasonable value for the drag coefficient. However, it is difficult to extend the model to a wake without a splitter plate that experiences vortex shedding (Roshko 1993; Williamson 1996*a*). Balachandar *et al.* (1997) conducted a thorough review of past wake models and presented detailed contours of flow properties in the wake of cylinders (circular and rectangular) as well as flat plates.

Correlations of drag versus flow speed for a given bluff-body shape are complicated by factors such as flow transitions, surface roughness effects, and the choosing of the proper characteristic length and area to be used in the Reynolds number and drag coefficient. Rotation effects present an added difficulty. For these reasons, no single correlation, empirical or otherwise, has been found to relate drag force to flow speed for different body shapes and for all flow speeds.

The goal of the present study is to arrive at a correlation between force and motion that is valid for various body shapes.

## 2. An Integral Quantity Characterizing Vortices

In many bluff-body flow problems, general characteristics of the velocity field are fairly well known, at least in the region most critical to drag prediction, namely the wake bubble. In fact, the common trait among bluff-body flows appears to be the existence of a wake bubble in the time-mean flow field. For these reasons, a drag correlation is proposed in the present study that involves an integral property of the mean wake vortex, that is, either of the symmetric time-mean vortices in the wake bubble.

Drag can be studied qualitatively by comparing the drag characteristics of two limiting extremes. Figure 1 shows a bluff body that experiences significant drag, and Figure 2 shows a knife edge in inviscid flow that experiences no drag. If a drag relation is to be obtained that is generally applicable, it should exhibit the proper behavior in the limiting case represented by Figure 2. Therefore, the two conceptual extremes of Figures 1 and 2 will help to reveal the functional form of the drag correlation.

The application of the drag correlation to follow (see (1)) will require a clear distinction between the definition of the fluid domain and that of the fluid boundary. The strict mathematical definition of a boundary would stipulate an infinitely thin set of points around the body. In contrast, Prandtl's boundary *layer* is of finite thickness and extends to the locus of points where the velocity has nearly reached its asymptotic (ideal flow) value. The boundary layer concept has proven to be extremely valuable in cases where it remains attached; however, in the context of bluff-body flows, it is obviously difficult to apply. In the present work, the term *boundary* will refer to neither of the above definitions, but rather to all fluid possessing vorticity. In bluff body flows at appreciable Reynolds numbers, this will coincide closely with the fluid enveloped by closed streamlines since comparatively little vorticity exists outside the wake bubble.

Consider Figure 1 to represent a stationary body in a wind tunnel in which force is transmitted from the wind tunnel walls to the body. The wake bubble is the effect of this force that has been transmitted from the stationary body to the boundary of the fluid, but which effect has *not yet entered the fluid domain*. The energy is still contained within the boundary, which, for Figure 1, is of finite thickness, i.e., the extent of the mean wake bubble. In this case, the thickness of the boundary, as defined here, is finite; the vorticity within the boundary is finite; and a drag force exists.

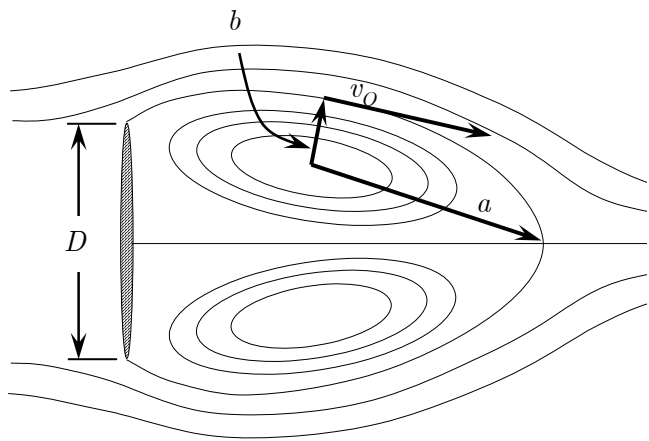

Figure 1. Schematic of flow past a normal flat plate.

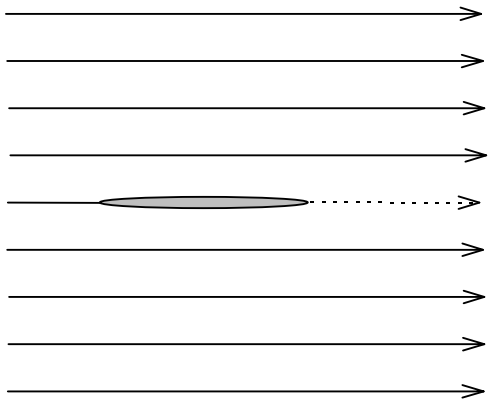

Figure 2. Schematic of flow past a knife edge.

If the wake size were to somehow decrease, with the velocity $v_O$ on its perimeter held constant (see Figure 1), the average vorticity in the boundary would obviously increase. As mentioned above, Figure 2 represents the limit of this wake reduction process, wherein the flow is inviscid and irrotational outside the now infinitely thin boundary. The distinction with previous definitions of the term "boundary" is that even in ideal flow, a "boundary," as defined here, still exists, albeit with only one streamline – the one that defines the body contour. Prandtl's boundary layer, on the other hand, does not exist in such a flow.

If a fluid boundary is to exist in this limiting case, the abrupt change in velocity at the fluid/body interface requires that the vorticity within that boundary be infinite. Therefore, the case of zero drag, i.e., ideal flow, corresponds to infinite vorticity in the fluid boundary. Skin friction drag and form drag are contained within the two extremes of Figures 1 and 2.

From a qualitative understanding of the drag in Figures 1 and 2, an infinite vorticity in the boundary of a fluid domain is seen to correspond to inviscid flow. Consequently, for a given free stream velocity, aerodynamic drag is inversely proportional to the vorticity within the boundary of the fluid. Part of the reason for this counterintuitive result is due to the redefinition of the term *boundary*, without which, the following parameter cannot be defined:

$$G = \int_{V} \frac{e_K}{\omega} dV. \tag{1}$$

In the definition of this parameter, which will be used in the drag correlation shown hereafter, $e_K$ is the kinetic energy of the fluid per unit volume, $\omega$ is the vorticity magnitude within the boundary volume, and $V$ is the volume of the fluid boundary as defined above. If the boundary volume is not defined as described above, then $G$ can become infinite if applied in regions external to the wake bubble, where $\omega = 0$. Neither Prandtl's boundary layer nor the mathematical definition of a boundary resolves this problem suitably. Flows containing only one closed streamline, namely that coinciding with the contour of the body itself in ideal flow, produce no drag due to the infinite vorticity in the boundary.

In Table 1, the quantity $G$ is evaluated inside various rotating, non-translating shapes. For the rigid bodies listed in Table 1, as well as the elliptical patch of uniform vorticity in planar flow, the following relation is found to hold:

$$G = \frac{1}{4}L, \tag{2}$$

| Config. | $\dot{m}$ | $\dfrac{\dot{m}}{m}$ | $G = \int \dfrac{e_K}{\omega} \, \mathrm{d}V$ | $L = \int |\boldsymbol{\ell}| \, \mathrm{d}V$ | $\dfrac{G}{L}$ |
|---|---|---|---|---|---|
| Elliptical vortex patch | $\dfrac{1}{2}\rho v_O b W$ | $\dfrac{v_O}{2\pi a}$ | $\dfrac{1}{8}\rho v_O \pi a b^2 W$ | $\dfrac{1}{2}\rho v_O \pi a b^2 W$ | $\dfrac{1}{4}$ |
| Solid Cylinder (radius $= b$) | $\dfrac{1}{2}\rho v_O b W$ | $\dfrac{\Omega}{2\pi}$ | $\dfrac{1}{8}\rho v_O \pi b^3 W$ | $\dfrac{1}{2}\rho v_O \pi b^3 W$ | $\dfrac{1}{4}$ |
| Solid Cone (base $= b$ height $= h$) | $\dfrac{1}{6}\rho \Omega b^2 h$ | $\dfrac{\Omega}{2\pi}$ | $\dfrac{1}{40}\rho \Omega \pi b^4 h$ | $\dfrac{1}{10}\rho \Omega \pi b^4 h$ | $\dfrac{1}{4}$ |
| Solid spinning Sphere | $\dfrac{2}{3}\rho \Omega R^3$ | $\dfrac{\Omega}{2\pi}$ | $\dfrac{2}{15}\rho \Omega \pi R^5$ | $\dfrac{8}{15}\rho \Omega \pi R^5$ | $\dfrac{1}{4}$ |
| Solid Torus ($R_C \geq r_O$) | $\rho A \Omega R_C$ | $\dfrac{\Omega}{2\pi}$ | $\dfrac{\rho \Omega A C}{4}\left[\dfrac{3}{4}r_O^2 + R_C^2\right]$ | $\rho \Omega A C \left[\dfrac{3}{4}r_O^2 + R_C^2\right]$ | $\dfrac{1}{4}$ |
| Hill's spherical vortex | $\dfrac{\pi}{4}\rho v_O R_O^2$ | $\dfrac{3 v_O}{16 R_O}$ | $\dfrac{\rho \pi^2 v_O R_O^4}{30}$ | $\dfrac{\rho \pi^2 v_O R_O^4}{8}$ | $\dfrac{4}{15}$ |

Table 1. Integral properties of rotating material in various shapes. ($W$ is the length along the symmetry axis, $\Omega = v_O/b = v_O/R_O = v/R$, rotation is about the symmetry axis, $\boldsymbol{\ell} = R \times \rho v$, $\rho$ and $m$ are the density and mass of the material, respectively. For the torus: $A = \pi r_O^2$ is the torus cross-section area, and $C = 2\pi R_C$ is the circumference of the core circle. For Hill's spherical vortex, $R_O$ is the sphere radius.)

where $L$ is the angular momentum of the body or material. Although $G$ and $L$ are both evaluated for the same volume of rotating material, the relation in (2) appears not to be a kinematic identity because, as seen from Table 1, it is not true for the core of Hill's spherical vortex. The following relation, however, does hold for the spherical vortex as well as the solid cylinder and the elliptical patch of uniform vorticity:

$$L = \dot{m} S, \qquad (3)$$

where $S$ is the cross-sectional area of the vortex containing a mean circulating flow rate of $\dot{m}$. The mass flow rate $\dot{m}$ is that passing through the area $A = bW$, where $W$ is the length of the vortex into the page (see Figure 1); therefore, $V = 2SW$ for a pair of identical vortices. Considering the complexity of the flow in the core of Hill's spherical vortex and the non-uniform distribution of mass flux along any axis from its core center to its perimeter, (3) is clearly not a trivial statement. The validity of (3) is naturally independent of the type of material enclosed.

Observe from Table 1 that for the elliptical patch of uniform vorticity, the number $G/S$ (as well as $G/V$) is independent of the ellipticity of the vortex cross section. Note that for all but the spherical vortex, the ratio $\dot{m}/m$ is the frequency of rotation. For the spherical vortex, the frequency is spatially dependent and goes to infinity on the bounding streamline due to the two stagnation points on that streamline.

## 3. Drag Correlation

As the wake bubble becomes fully turbulent with increasing Reynolds number, drag correlations involving the quantity $\dot{m}S$ in (3) appear to break down, while those involving the quantity on the left side of (2) remain valid (see Figure 3). According to the far right column in Table 1, (2) cannot be assumed valid generally. However, in the present study, only two-dimensional bodies are considered; therefore, the spherical vortex is not immediately relevant.

The experimental data in Figure 3 suggest a relationship of the following form:

$$c_D\,Re \;=\; k\frac{G}{\mu\forall}, \qquad (4)$$

where $\mu$ is the fluid viscosity, and $\forall$ is the volume of either of the elliptical vortices in the time-mean wake bubble. The quantity $c_D Re$ will be referred to as the *dimensionless drag*. Figure 3 shows the correlation of dimensionless drag with $G/(\mu\forall)$ for available whole-field wake data in nominally two-dimensional flows (listed in Table 2). The integration in (1) is carried out over the time-mean wake vortex, i.e., the fluid boundary. For this integration, the time-mean wake vortices trailing the symmetric bluff bodies were modeled by the elliptical patch of uniform vorticity, as explained in §4. Geometrical parameters of the pair of elliptical vortices, such as $b/D$ (defined by Figure 1), were taken from available bluff-body wake measurements, cited in the left-most column of Table 2.

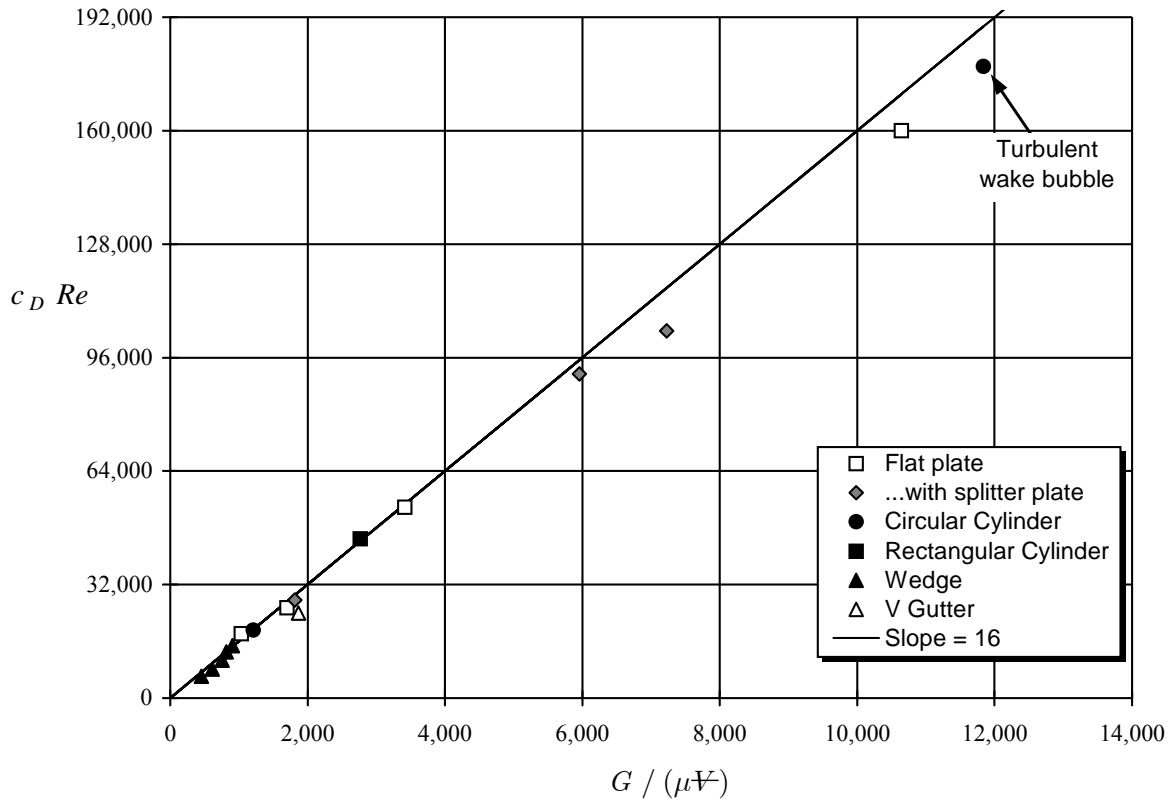

Figure 3. Variation of $c_D Re$ with $G/(\mu\forall)$ for various obstacles. Data references are tabulated in Table 2.

| Study | Geometry | $Re$ | $\beta$ | $AR$ | $St$ | $a/D$ | $b/D$ | $v_O/v$ | $c_{D\text{ Pred}}$ | $c_{D\text{ Data}}$ |
|---|---|---|---|---|---|---|---|---|---|---|
| Arie & Rouse (1956) | Normal plate & splitter plate | 75,000 | 8.3% | 12 | | | 0.47 | 0.82 | 1.54 | $1.38^a$ |
| Castro & Haque (1987) | Normal plate & splitter plate | 23,000 | 6.5% | 12.4 | | | 0.46 | 0.79 | 1.45 | $1.38^b$ |
| Ruderich & Fernholz (1986) | Normal plate & splitter plate$^c$ | 32,000 | 10% | 22 | | | $0.6^d$ | 1.15 | 2.75 | n.a. |
| Good & Joubert (1968) | Normal fence in boundary layer | 63,000 | 11% | 11 | | | 0.72 | 1.05 | $1.51^e$ | $1.45^f$ |
| Perry & Steiner (1987) | Flat plate | 20,000 | 24.6% | 9.4 | 0.17 | 1.1 | 0.29 | | 1.36 | $1.27^g$ |
| Leder & Geropp (1993) | Flat plate | 28,000 | 8.5% | n.a. | 0.14 | 1.32 | 0.42 | | 1.95 | 1.92 |
| Fail et al. (1957) | Flat plate | 86,000 | 2.6% | 29 | 0.145 | 1.6 | 0.33 | 0.75 | 1.92 | $1.86^h$ |
| Bradbury (1976) | Flat plate$^i$ | 25,000 | 10% | 15 | | | 0.36 | 0.9 | 2.59 | n.a. |
| Bachalo et al. (1993) | Cylinder | 16,000 | 16.4% | 6.1 | 0.19 | 1.15 | 0.22 | | 1.21 | 1.20 |
| Cantwell & Coles (1983) | Cylinder | 144,000 | 4% | 29 | 0.179 | 0.43 | 0.12 | | 1.31 | 1.237 |
| Lyn et al. (1995) | Rectangular Cylinder | 21,400 | 7% | | 0.132 | | $0.34^j$ | 0.76 | 2.07 | 2.1 |
| Okamoto et al. (1977) | Wedge: | | | | | | | | | |
| | $\alpha = 15$ | 9,000 | 2.5% | 40 | 0.245 | 0.525 | 0.25 | | 0.81 | 0.68 |
| | $\alpha = 30$ | 9,000 | 2.5% | 40 | 0.24 | 0.69 | 0.26 | | 1.08 | 0.91 |
| | $\alpha = 60$ | 9,000 | 2.5% | 40 | 0.214 | 0.88 | 0.285 | | 1.35 | 1.19 |
| | $\alpha = 90$ | 9,000 | 2.5% | 40 | 0.179 | 1.01 | 0.32 | | 1.45 | 1.45 |
| | $\alpha = 120$ | 9,000 | 2.5% | 40 | 0.164 | 1.1 | 0.355 | | 1.61 | 1.64 |
| | $\alpha = 180$ | 9,000 | 2.5% | 40 | 0.143 | 1.25 | 0.41 | | 1.84 | 2.01 |
| Yang & Tsai (1992) | V Gutter (30°) | 24,000 | 20% | 5 | $0.225^k$ | 0.55 | 0.4 | | 1.24 | $1.0^l$ |

Table 2. Comparison between the drag coefficient of various two-dimensional bodies and the values predicted by (22). $Re > 9,000$.

---

[a] This value obtained by Arie & Rouse omits the drag of their long ($10D$) splitter plate by virtue of the location at which they measured the downstream pressures in their drag integral.

[b] This value by Arie & Rouse (1956) omits the drag of the splitter plate ($17D$ in this case).

[c] Length of splitter plate $= 30D$. Ruderich & Fernholz intentionally tested with very high free stream turbulence (~30%).

[d] According to Figure 8 of Castro & Haque (1987), velocity profiles appear to cross at the common value of $v_O$ at the vertical distance $b$ from vortex center to the wake bubble boundary. This fact was used to determine $b$ from Figures 15 of Ruderich & Fernholz (1986). Figure 16 was then use to find $v_O$.

[e] The correlation value was divided in half because only one vortex was present.

[f] From Figure 6 of Good & Joubert (1968) with $h/\delta = 2.34$.

[g] Taken from Fail *et al.* (1957) data for appropriate aspect ratio.

[h] Their value of 1.96 based on Fage & Johansen (1927) was corrected for blockage by Fail *et al.* (1957) using Maskell's method (see Maskell 1965).

[i] Bradbury intentionally tested with very high free stream turbulence. Vortex shedding is assumed here ($\delta = 1$, see §5), though no Strouhal number or drag coefficient was reported in the study.

[j] Can be found by locating in Figure 3 of Lyn *et al.* (1995) where $\overline{v}^t$ reaches a minimum and subtracting from the location where the streamwise velocity, $\overline{u}^t$, is zero, also in Figure 3 of the paper.

[k] Yang *et al.* (1994).

[l] Hoerner (1965) p. 3-18.

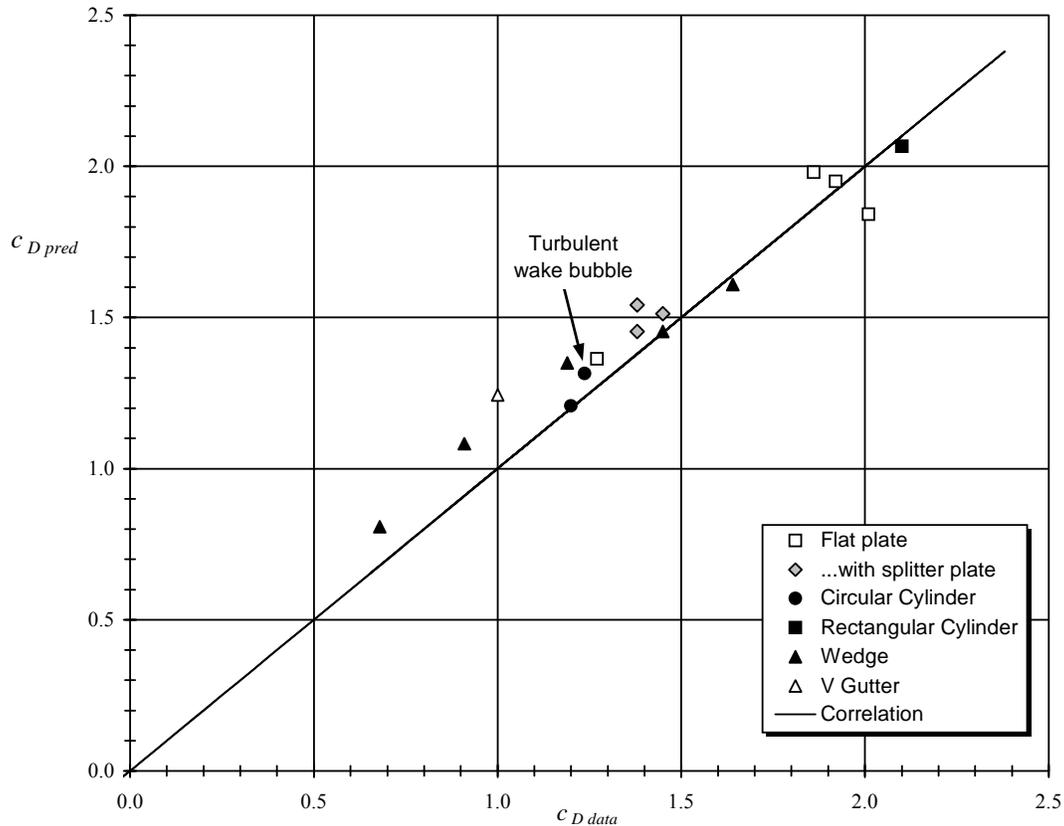

Figure 4. Correlation of predicted values of $c_D$ with data listed in Table 2.

The far right column in Table 2, labeled "$c_{D\;data}$," shows values of drag coefficient obtained from the experimental studies listed in the left column of the table or in its footnotes. As seen from Figure 3, the data suggest that $k \approx 16$ in (4). The column labeled "$c_{D\;pred}$" in Table 2 shows values predicted by (4) with $k = 16$. (The actual equation used is (22), which is obtained by solving for $c_D$.)

The aspect ratio, $AR$, shown in Table 2 is defined as the ratio of the spanwise body dimension to its transverse dimension, and the blockage, $\beta$, is defined as the ratio of the transverse body dimension to the wind tunnel height measured in the same direction. All test bodies spanned the test sections or used end plates. Figure 4 shows the correlation of predicted and measured drag coefficient.

The correlation in (4) predicts a reduction in drag with increasing vorticity in the wake, as dictated by Figure 2. It also suggests that a reduction in drag may be possible by increasing the vorticity in the wake of the body, such as by spinning the body appropriately. Such spinning of, for example, a disc upon re-entry into the atmosphere would also distribute the stagnation point along the entire perimeter of the body rather than confine it to a single point, thereby lowering the overall peak temperature. For the case of supersonic flight of such a spinning body, it is not yet known whether the sonic boom would be reduced relative to the non-spinning case.

Equation (4) may be cast in the form:

$$F_D = \frac{kG}{2S} v_\infty, \qquad (5)$$

where $F_D$ is the drag force, and $v_\infty$ is the free stream velocity.

The form of the drag correlation can also be obtained by using Newton's law, $F = m\dot{v}$, as a form for the correlation, assuming that the mass of fluid trapped in the wake vortex is constant. This allows a force to be associated with the mass, $m$, rotating in the wake vortex (assumed circular for the moment) having a radial acceleration given by $\dot{v} = \Omega v$. The drag correlation therefore becomes:

$$F_D \sim m\,\Omega\,v.$$

Here, $m$ can be replaced with $m = \dot{m}/f$ according to Table 1, where the frequency $f$ is given by $\Omega/(2\pi)$. Making this substitution gives:

$$F_D \sim 2\pi \dot{m} v. \qquad (6)$$

From (3), this is equivalent to:

$$F_D \sim 2\pi \frac{L}{S} v,$$

and by (2), we have the form for the drag correlation:

$$F_D \sim 8\pi \frac{G}{S} v. \qquad (7)$$

Compare the relation in (7) with (5). Correlating bluff-body drag with the radial acceleration in the wake vortex, as done here to obtain (7), might have been anticipated if one were to argue that the centrifugal force is acting against the back side of the body where the wake vortex is located.

## 4. Evaluating $G$ in a Planar Wake Vortex

Equation (4) or (5), together with the calculation of $G$ to be described in §4.2, will relate the drag coefficient of two-dimensional bodies at high Reynolds numbers to an integral property of their wakes, expressed in terms of the velocity at a point on the perimeter of the time-mean wake vortex and a dimension of this vortex. The mean velocity field can then be determined from the analytical solution for the flow field in the wake bubble. As observed by Roshko (1967), even for flows exhibiting vortex shedding, the time average in the near wake remains a pair of counter-rotating vortices (see also Perry and Steiner 1987). These are known experimentally to have a linear velocity profile along their semi-minor axes (see, e.g., Castro and Haque 1987, Yang and Tsai 1992). If this fact, along with the continuity equation, is applied to a stationary elliptical vortex, the resulting vorticity field is uniform (see Morton 1997). Therefore, for the nominally two-dimensional flows considered herein, the time-mean velocity field of the wake bubble can be approximated quite well using a pair of stationary elliptical patches of uniform vorticity. The velocity field in an elliptical patch of uniform vorticity can be found by constructing a streamlined coordinate system made of concentric elliptical streamlines and applying the continuity equation inside (see Morton 1997; also summarized in §4.1 below).

### 4.1. Velocity, Pressure, and Vorticity Fields in the Time-Mean Wake Vortex

A streamline coordinate system can be defined in a time-mean wake vortex as follows:

$$x = \bar{x}^1 A \cos(\bar{x}^2) \qquad y = \bar{x}^1 B \sin(\bar{x}^2) \qquad z = \bar{x}^3, \tag{8}$$

where $x$, $y$, and $z$ are coordinates in a Cartesian system, and different values of the coordinate $\bar{x}^1$ correspond to different elliptical streamlines. Application of the continuity equation requires that the streamline component of the velocity tensor, $d\bar{x}^k/dt$, be (see Morton 1997):

$$\bar{v}^2 = \frac{v_O}{a}, \tag{9}$$

where $v_O = \Omega_O b$, the lengths $a$ and $b$ are, respectively, the semi-major and semi-minor axes of the elliptical vortex, and $\Omega_O$ is the rotation rate about the vortex center of a fluid particle at the point where the semi-minor axis intersects the perimeter of the vortex. Transforming from streamline coordinates to rectangular coordinates gives:

$$\boldsymbol{v} = \left[-y, \ \left(\frac{b}{a}\right)^2 x, \ 0\right] \Omega_O. \tag{10}$$

The pressure field in the vortex is (see Morton 1997):

$$p - p_C = \frac{1}{2}\rho(\Omega_O r)^2 \left(\frac{b}{a}\right)^2, \tag{11}$$

where $p$ is the pressure at a distance $r$ from the vortex center where the pressure is $p_C$. Therefore, pressure contours are circular, regardless of the ellipticity of the vortex cross section. The constant vorticity of the elliptical patch is (see Morton 1997)

$$\omega = \bar{v}^2 \left(\frac{a}{b} + \frac{b}{a}\right), \tag{12}$$

or

$$\omega = \frac{v_O}{b}\left[1 + \left(\frac{b}{a}\right)^2\right]. \tag{13}$$

### 4.2. Evaluating $G$

The ratio $G$ to be used in (5) can be found by adding the turbulent kinetic energy per unit volume, $e_{TK}$, and the mean kinetic energy, $e_{MK}$, as follows:

$$G \equiv \int_{\mathcal{V}} \frac{e_K}{\omega} d\mathcal{V} = \int_{\mathcal{V}} \left(\frac{e_{MK} + e_{TK}}{\omega}\right) d\mathcal{V}. \tag{14}$$

The standing eddies in symmetric bluff-body wakes naturally occur in pairs, and if these are considered to be a single vortex ring infinitely long into and out of the page, the kinetic energy is doubled, but the vorticity is not. For a pair of identical vortices, therefore,

$$G = 2\int_{\mathcal{V}} \left(\frac{e_{MK} + e_{TK}}{\omega}\right) d\mathcal{V}. \tag{15}$$

One need not consider $\omega$ to be the absolute value of vorticity in order to account of the vorticity of opposite signs in the wake, provided one considers the volume of each half of the wake to be oriented in the direction of its vorticity. Expanding (15) gives

$$G = \int_0^W \int_0^{2\pi} \int_0^{\bar{x}^1_{max}} \frac{\bar{g}_{22}\left(\bar{v}^2\right)^2}{\sqrt{\bar{g}_{33}}\bar{\omega}^3} \rho\sqrt{\bar{g}}\,d\bar{x}^1 d\bar{x}^2 d\bar{x}^3 \; + \; \int_{\text{eddy}} \frac{\left(\langle v_1'^2\rangle + \langle v_2'^2\rangle + \langle v_3'^2\rangle\right)}{\omega} \rho\,d\mathcal{V}, \quad (16)$$

where, for the streamlined coordinate system defined by (8), $\bar{g}_{33} = 1$, $\sqrt{\bar{g}} = \bar{x}^1 A B$, $\bar{\omega}^3$ is the constant given by (12) or (13), and $\bar{g}_{22} = (\bar{x}^1)^2[A^2\sin^2(\bar{x}^2) + B^2\cos^2(\bar{x}^2)]$. The velocity components marked with primes in (16) indicate the random fluctuations of the respective components, and the angled brackets indicate averaging. According to (9), $\bar{v}^2$ is constant and, like $\bar{\omega}^3$, can be brought out of the integral. Performing the integration in (16) for the pair of vortices gives:

$$G = \frac{1}{4}\rho v_O b \mathcal{V} + \frac{\rho v_\infty^2}{\omega} k_T \mathcal{V}, \quad (17)$$

where $\mathcal{V}$ is the vortex volume and

$$k_T \equiv \frac{1}{\mathcal{V}}\sum_i \left[\frac{\langle v_1'^2\rangle}{v_\infty^2} + \frac{\langle v_2'^2\rangle}{v_\infty^2} + \frac{\langle v_3'^2\rangle}{v_\infty^2}\right]\Delta\mathcal{V}_i. \quad (18)$$

Notice from (17) that when turbulent kinetic energy is negligible in the wake vortex, $G/S$ is independent of the axis ratio of the elliptical vortex and that, according to (2) and (3), it is equal to half the mass flow rate through any semi-axes of the pair of vortices. That is:

$$\frac{G_M}{S} = \frac{\dot{m}}{2}. \qquad \text{(planar, laminar)} \quad (19)$$

Therefore the ratio $G/S$ is invariant with respect to the axis ratio of the vortex cross section, for a given semi-minor axis and perimeter velocity $v_O$.

Therefore, in the case of laminar wake bubbles, (2) and (3) can be substituted into (4) to see that the drag coefficient is proportional to $\dot{m}/\dot{m}_\infty$, where $\dot{m}$ is the flow rate of mass recirculating in the wake bubble, and $\dot{m}_\infty$ is the flow rate of mass that would otherwise pass, with velocity $v_\infty$, through the projected area made by the semi-minor axis of the wake vortex and its length.

## 5. Drag Coefficient

The drag coefficient can be computed by substituting (13) into (17), and the result into (4), giving:

$$c_D = 8\frac{b}{D}\left[\frac{v_O}{2v_\infty} + \frac{2v_\infty}{v_O}\frac{k_T}{\left(1 + (b/a)^2\right)}\right]. \quad (20)$$

In all of the cases studied herein, namely two-dimensional flows, the characteristic length $D$ is the projected body dimension normal to the free stream and is used in the Reynolds number. In all values of Table 2, except for those of Cantwell and Coles (1983) (which will be discussed in §6 and §7), the turbulent transition waves (Bloor 1964) had not penetrated into the wake vortex so that $k_T = 0$ in those cases. If the velocity ratio appearing in (20) is not known, it can often be related to the shedding frequency by a Strouhal number grouping derived from the elliptical vortex patch (see Appendix). The necessary relation from that derivation is repeated below for convenience:

$$\frac{v_O}{v_\infty} = \frac{2\pi a}{nD} St \,. \tag{21}$$

Here, $St$ is the classical definition of the Strouhal number, and $n$ is the number of vortices shed downstream per shedding cycle, namely 2. Equation (21) can be employed to write (20) in terms of the Strouhal number rather than the vortex perimeter velocity ratio. With the onset of vortex shedding, however, the energy becomes split into a mean component and a periodic component. The mean wake bubble dramatically decreases in size with the removal of a splitter plate. This is clearly demonstrated by Hwang *et al.* (2003) at lower Reynolds number. If the correlation for the drag coefficient, (20), is modified so as to fit the data in Table 2 with and without splitter plates, the following form is obtained:

$$c_D = 8\frac{b}{D}\left[(1+\delta)\frac{v_O}{2v_\infty} + \frac{2\,v_\infty}{v_O}\frac{k_T}{\left(1+(b/a)^2\right)}\right]. \tag{22}$$

When vortex shedding is absent due to the presence of a splitter plate and the attendant large wake bubble, $\delta = 0$; and when vortex shedding is present and the mean wake bubble shrinks, $\delta = 1$. Therefore, for a non-turbulent wake bubble, kinetic energy of the mean flow is cut in half when the splitter plate is removed, the missing half being in the form of vortex shedding. Since all non-turbulent kinetic energy is accounted for (in the present work) by considering only the time-mean wake bubble, the kinetic energy due to vortex shedding must be included separately through the use of $\delta$.

Using (22), drag predictions can be made from steady or time-mean wake flow visualizations; however, the perimeter velocity, $v_O$, must be measured experimentally. The dimensionless distance, $b/D$, to this point from the vortex center may be found graphically from visualization of the mean flow (see Figure 1). Alternatively, by using (21), the measured Strouhal number and dimensionless semi-major axis, $a/D$, of the wake vortex may be used in lieu of the velocity ratio.

## 6. Illustrative Example

The case to be examined is the set of cylinder experiments made by Cantwell and Coles (1983). The purpose of their experiment was to provide measurements of wake flow in which the separation point was laminar, yet the near wake region was fully turbulent. As it turns out, this strategy has provided an invaluable set of data, which allows (1) or (4) to be seen in a more general light. Perrin *et al.* (2006) made a similar comprehensive study at essentially the same Reynolds number but with a low aspect ratio cylinder and higher blockage. In accordance with their objective stated above, Cantwell and Coles chose from a series of cases, one in which the Reynolds number was $1.44 \times 10^5$ so that the flow was on the brink of entering the so-called drag crisis.

| $Re$ | $c_D$ | $St$ |
|---|---|---|
| $2 \times 10^5$ | 1.14 | 0.19 |
| $4.1 \times 10^5$ | 0.235 | 0.46 |

Table 3. Values of drag coefficient and Strouhal number before and after the drag crisis. (Bearman 1969)

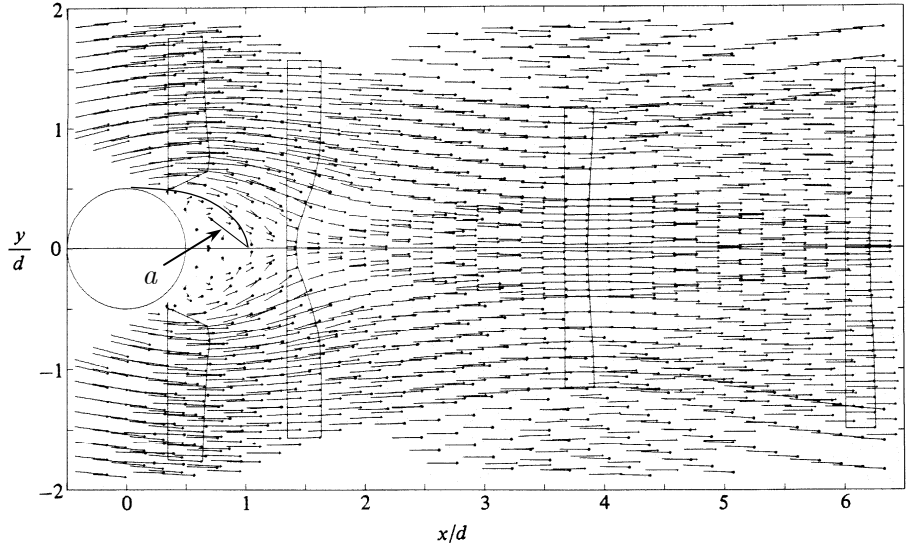

Figure 5. Mean velocity vectors in the wake of a circular cylinder. $Re = 144{,}000$, $AR = 25$, $\beta = 4\%$, $c_D = 1.237$, $St = 0.179$, $TL = 0.6\%$. (adapted from Cantwell and Coles 1983)

Table 3 shows values reported by Bearman (1969) for the Strouhal number and drag coefficient on either side of this drag crisis. The mean velocity measurements of Cantwell and Coles (1983) immediately below the flow speed at which this drag crisis occurs are shown in Figure 5 with $a$ indicated. Since the velocity at the semi-minor axis of the wake vortex was not given, (21) will be used to estimate it.

At this Reynolds number, the turbulent transition waves have penetrated the wake bubble. The result is a noticeable reduction in the size of the wake bubble and a dramatic increase in the *fluctuating* component of velocity. Measurements of the needed velocity fluctuations were also made by Cantwell and Coles. The turbulent kinetic energy in the wake bubble can be found using mean contours of the random fluctuations $\langle v_1'^2 \rangle / v_\infty^2$ and $\langle v_2'^2 \rangle / v_\infty^2$, shown (with overbars) in Figures 27(b) and 28(b) of their paper. The remaining quantity, $\langle v_3'^2 \rangle / v_\infty^2$, will be assumed to be an average of these two. The Reynolds normal stress contours reported by Cantwell and Coles (1983) appear very similar to those presented by Perrin *et al.* (2006) for a low aspect ratio cylinder and by Balachandar *et al.* (1997) for a cylinder wake bubble at low Reynolds number. From these and Figure 5 above, the semi-axes of the vortices are found graphically to be $a / D \approx 0.43$ and $b / D \approx 0.12$. Applying (18) graphically to Figures 27(b) and 28(b) of Cantwell and Coles (1983) shows that $k_T = 0.147$.

The velocity ratio $v_O / v_\infty$ is found using (21):

$$\frac{v_O}{v_\infty} = \frac{2\pi(0.43)}{2} 0.179 = 0.242.$$

The drag coefficient can then be estimated by (22), as follows:

$$c_D = 8\frac{b}{D}\left[(1+\delta)\frac{v_O}{2v_\infty} + \frac{2v_\infty}{v_O}\frac{k_T}{\left(1 + (b/a)^2\right)}\right]$$

$$c_D = 8(0.12)\left[(1+1)\frac{0.242}{2} + \frac{2}{0.242}\frac{(0.147)}{\left(1+(0.279)^2\right)}\right]$$

$$c_D \approx 1.31$$

The value measured by Cantwell and Coles was $c_D = 1.237$.

## 7. Discussion
### 7.1. The Drag Crisis

In the example shown, the drag coefficient remains large even after the increase in the turbulent kinetic energy within the wake bubble and the decrease in mean wake width. The drag coefficient finally decreases suddenly when the Reynolds number is increased still further into the range $2 \times 10^5 < Re < 5 \times 10^5$, known as the "drag crisis," due to the well-known shift in the location of the separation point. This shift causes the circulation (effectively the denominator of (5)) to increase suddenly. Recall that the Reynolds number in the Cantwell and Coles experiment was large enough for the mean wake bubble to become almost entirely turbulent, yet small enough to ensure that the separation point remain forward on the body. The range of Reynolds numbers for which this set of circumstances exists is narrow. At these unique conditions, the size of the wake bubble is much smaller than it is at slightly lower Reynolds numbers, for which the wake bubble is not fully turbulent. Therefore, as the Reynolds number is increased into this unique range, the increase in fluctuating kinetic energy in the wake bubble and the reduction in the size of the mean wake bubble occur simultaneously, with no noticeable change in the drag coefficient. The circulation in the wake bubble, on the other hand, has not changed appreciably within this narrow range of Reynolds numbers because circulation within the wake volume is a direct function of the distance traveled by the fluid through the boundary layer. The sudden reduction in the drag coefficient does not occur until the boundary layer extends to the rear of the cylinder, at an even higher Reynolds number. The distance traveled by the fluid in the boundary layer is then longer, the circulation in the denominator of the drag relation is therefore higher, and hence the drag coefficient suddenly decreases (see (4)). Therefore, the decrease in drag coefficient occurs at a slightly higher Reynolds number than does the decrease in wake size. Consequently, near the drag crisis, the reciprocal of circulation appears to be a much stronger predictor of drag than the mean wake size.

Situations do exist where advancement of the transition point forward on the body has the effect of reducing drag (see e.g., Coustols 1996, Gad-el-Hak & Bushnell 1991, Coustols and Schmitt 1990). However, this is accomplished by placing distributed roughness elements or riblets on the surface, which can themselves be viewed from the standpoint of increasing the circulation in the fluid boundary and hence reducing the drag.

### 7.2. Turbulent Angular Momentum

For turbulent flows, the drag correlation is easier to evaluate using the left side of (2) than the right. However, if the right side of (2) is substituted into (5), one obtains:

$$F_D = 2v_\infty \frac{L}{S}, \tag{23}$$

for each half of the wake bubble. Therefore, for a given velocity, a higher force causes a higher angular momentum per unit area of wake cross section. Turbulence is an ideal mechanism for filling a region with concentrations of angular momentum to accommodate a large force. From Table 1 and (3), we can write

$$\dot{m} = \frac{L}{S} = \frac{1}{2}\rho v_o b W \tag{24}$$

for a cylindrical vortex, the cross section of which is depicted in Figure 6$a$. If this cylindrical vortex shown were replaced with two counter-rotating vortices as shown in Figure 6$b$ while maintaining the same velocity magnitude at points of tangency, the length $b$ would become half of what it was in Figure 6$a$, as shown. There would be no gain in the storage capacity of angular momentum per unit area, $L/S$, (or mass flow rate) because, while there are now twice as many vortices, the quantity $L/S$, which according to (24) scales with $b$, is cut in half for each vortex. If, however, the original vortex were replaced with four vortices as shown in Figure 6$c$, there would be a gain in storage capacity of mass flow rate because $b$ in this case reduces to $b/(1+\sqrt{2})$, and consequently the mass flow rate $\dot{m}_i$ circulating in a single vortex decreases by the same ratio, $1/(1+\sqrt{2})$. Yet there are now $N = 4$ vortices. Therefore, the total storage capacity of mass flow rate increases by a factor of $4/(1+\sqrt{2})$, since $\dot{m} = \sum \dot{m}_i = N\dot{m}_i$.

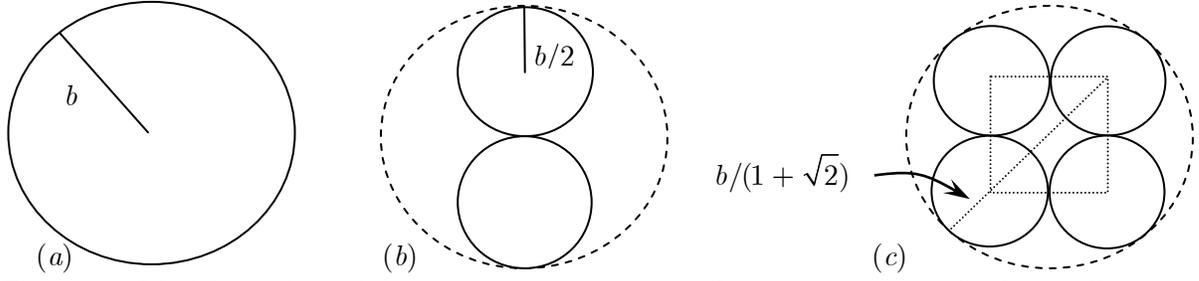

Figure 6. Model of the subdivision process of a single vortex into multiple smaller vortices. ($a$) $N = 1$, ($b$) $N = 2$, ($c$) $N = 4$.

In the limit as the number of vortex cross sections filling a given area increases, that is, as $N \to \infty$, we find by equating the total vortex cross-sectional areas before and after this subdivision process that the radius, and hence the mass flow rate $\dot{m}_i$ of each small vortex, scales as $\dot{m}_i \sim 1/\sqrt{N}$, so that the total storage capacity of mass flow rate within the volume scales as $\dot{m} \sim N\dot{m}_i \sim \sqrt{N}$. Therefore, according to these very rough order-of-magnitude arguments, the total mass flow rate enclosed in the wake volume can increase without bound as the vortices become smaller.

If the initial region were to subdivide into spherical rather than tubular vortices, the mass flow rate within each small vortex would scale as $\dot{m}_i \sim R^2$ according to Table 1. Equating the volumes before and after the subdivision process shows that the radius of each smaller spherical vortex scales as $R \sim 1/N^{1/3}$. Therefore, the mass flow rate within each of these vortices scales as $\dot{m}_i \sim R^2 \sim 1/N^{2/3}$; so that the total storage capacity of mass flow rate in the volume scales as $\dot{m} \sim N\dot{m}_i \sim N^{1/3}$. Therefore, again, the mass flow rate increases without bound as the number of vortices in a given region increases.

In general, therefore,

$$R \sim \frac{1}{N^{1/n}}, \tag{25}$$

$$\dot{m} = \sum_{i=1}^{N} \dot{m}_i = N\dot{m}_i \sim N^{1/n}, \tag{26}$$

where $n$ is the dimensionality of the vortex, and $R$ is the radius of the vortex, whether tubular ($n = 2$) or spherical ($n = 3$). Therefore, even though $N$ is greater for a packing of spherical vortices than for a packing of cylindrical vortex tubes of equal radius, as shown in the preceding two paragraphs, the mass flow rate in each of the $N$ vortex *tubes* is likely greater than it is in each of the $N$ *spherical* vortices, the general relation being: $\dot{m}_i \sim 1/N^{(n-1)/n}$. Substituting (25) into (26) shows that the two effects exactly compensate one another so that the total mass flow rate in the wake increases as $1/R$, where $R$ is the radius of the small vortices, whether tubular or spherical.

In §3, it was stated that the drag correlation using $\dot{m}$ failed when applied to highly turbulent wake bubbles, yet the correlation using $G$, which employs the ratio of kinetic energy to circulation, remained valid (recall there is no distinction in these two for a laminar vortex). This is because calculating $\dot{m}$ for a turbulent wake bubble based upon Figure 6*a* is inaccurate (and counting turbulent eddies in the equivalent of Figure 6*c* is impractical).

In summary, substitution of (3) into (23) gives, for both halves of the wake bubble:

$$F_D = 4\dot{m}v_\infty, \qquad (27)$$

where $\dot{m}$ is the total mass flow rate circulating in eddies in the fluid boundary. If a relation such as (27) or (5) should turn out to be valid generally, not only would it relate drag to flow speed without the use of viscosity, but, in the form of (27), it may be useful in turbulence modeling or as an eddy size criterion in large eddy simulation studies.

In turbulent flows, the quantity $\dot{m}$ cannot be computed reliably by assuming a flux across a semi-axis of wake vortices. If the wake bubble is so chaotic that $\dot{m}$ cannot be estimated, the ratio of kinetic energy to circulation contained in (1) provides a reasonably measureable alternative quantity which seems to serve the same purpose as $\dot{m}$.

### 7.3. Extending the Correlation to Lower Reynolds Numbers

At lower Reynolds numbers, sufficient whole-field data does not exist to formulate a meaningful correlation, owing to the complexity of the transitions in this regime. However, some of the most complete (for the purposes of the present study) works to date in the range $100 \leq Re \leq 300$, are listed in Table 4. For a description of the changes occurring in the wake near these Reynolds numbers, see Williamson (1996*a* or 1996*b*) for the circular cylinder and Julien *et al.* (2003) for the flat plate.

In order to better understand the three-dimensional behavior of flow in the near wake of a normal flat plate, Wu *et al.* (2005) analyzed the phase difference of vortex shedding at two different spanwise locations and found two distinct modes and very pronounced three-dimensionality. These two distinct shedding modes, also reported by Najjar and Balachandar (1998), may be related to the effect originally reported by Fail *et al.* (1957). In particular, Fail *et al.* noted that there are generally two shedding frequencies for each plate, "one associated with the smaller dimension of the plate, and a lower frequency associated with the larger dimension."

Note from Table 4 that just beyond the 3D wake transition regime where changes begin to occur in the nature of out-of-plane shedding, the correlation underpredicts the drag coefficient by as much as a factor of 2. Beyond $Re \sim O(10^3)$ on the other hand, explicitly accounting for 3D fluctuations is unnecessary, as evidenced by Figures 3 and 4. This may indicate that at these higher Reynolds numbers, any three-dimensional contribution remains a constant fraction of the total as flow speed changes, so that the correlation need only consider that in the nominal flow plane.

| **Study** | **Geometry** | $Re$ | $St$ | $a/D$ | $b/D$ | $v_O/v$ | $c_{D\,\text{Pred}}$ | $c_{D\,\text{Data}}$ |
|---|---|---|---|---|---|---|---|---|
| Najjar & Balachandar (1998) | Flat plate (3D DNS) | 250 | 0.167 | 1.0 | 0.28 | | 1.17 | 2.36 |
| Dennis & Chang (1970) | Cylinder (2D Steady) | 100 | - | 4.5 | 0.33 | n.a. | | 1.34 |
| Hwang, *et al.* (2003) | Cylinder (2D Unsteady) | 100 | 0.164 | 0.92 | 0.23 | | 0.87 | 1.33 |
| Hwang, *et al.* (2003) | Cylinder & Slip Plate (2D Unsteady) | 100 | 0.137 | 1.8 | 0.28 | n.a. | 0.87 | 1.18 |
| Mittal (2003) | Cylinder & Slip Plate (2D Steady[a]) | 100 | - | | 0.32 | n.a. | | 1.065 |
| Protas & Wesfreid (2002) | Cylinder & Slip Plate (2D Steady[a]) | 150 | - | | 0.33 | 0.54 | 0.71 | 0.86 |
| Balachandar *et al.* (1997) | Cylinder (3D DNS) | 300 | 0.207 | 0.5 | 0.24 | | 0.61 | 1.27[b] |

Table 4. Comparison between the drag coefficient of various two-dimensional bodies and the values predicted by (22). $100 \leq Re \leq 300$, $\beta = 0\%$, $AR = \infty$.

### 7.4. Analogies with Electromagnetism

The analogies between electromagnetic theory and fluid mechanics are numerous and well known (see, e.g., Milne-Thomson 1968 p. 574). Hsu and Fattahi (1976), for example, found the interesting fact that a tornado cannot be formed if the ground plane is removed. Therefore, the touchdown of a tornado resembles the arcing of an electric spark such as lightening. In light of (2) and (5), a few additional observations can be made. First, there are some interesting parallels between the quantity $G$ of vortices and Planck's constant, $h$. Both are ratios of energy to frequency (since vorticity in (13) can be related to frequency using $\Omega_O = 2\pi f$), and both have units of angular momentum (see Bohr 1913). It is well known that if Planck's constant were zero, there would be no quantum mechanics, a value of zero for $h$ corresponding to classical mechanics. A value of zero for $G$ corresponds to inviscid flow, for which the commutator of the velocity gradient, namely the vorticity, is zero everywhere except on infinitely thin boundaries. Fluid resistance becomes appreciable as separation vortices appear, and in metals, superconductivity is lost as soon as vortices appear (Abrikosov 1988, p. 456). Also, by multiplying (27) by velocity in order to obtain the power lost due to drag, one obtains a relation that resembles the $I^2R$ losses in electrical power lines, the role of resistance being played by $\dot{m}$ in the fluid eddies. Finally, there are similarities between the free-stream velocity vector in bluff-body flow and the Poynting vector of an electromagnetic wave. Figure 7 illustrates the relation between the force, the net vorticity vector, and the fluid velocity of a two-dimensional bluff body during vortex shedding. This figure also illustrates why the period of experimentally measured drag oscillation is typically half that of the lift oscillations. Drag occurs with the shedding of either of the two vortices, whereas lift only occurs with the shedding of one of them. The three quantities in the cylinder wake depicted in Figure 7 resemble those of a plane-polarized electromagnetic wave. From the wake measurements of Yun *et al.* (2006, Fig. 10*c*), the wake of a sphere appears to be circularly polarized.

---

[a] Unperturbed solution.
[b] Wieselsberger (1921, 1922) obtained 1.22.

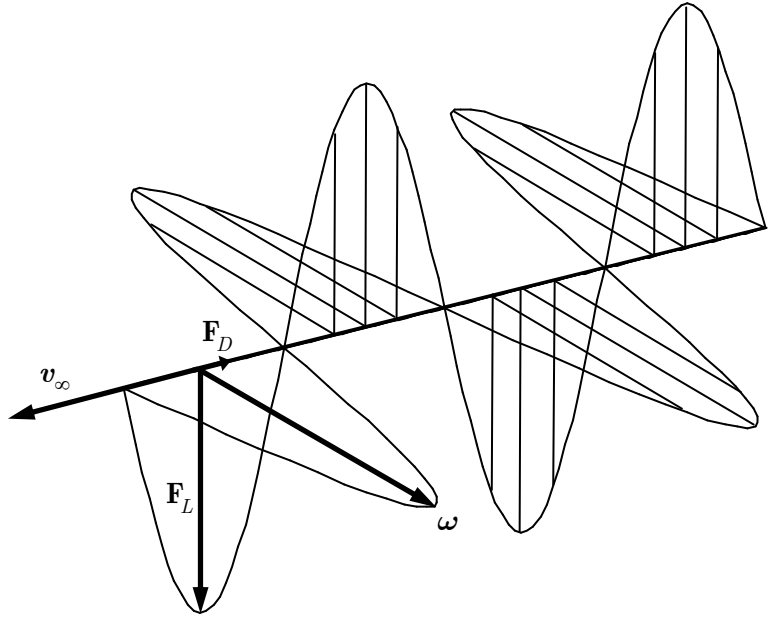

Figure 7. Plane-polarized quantities $\mathbf{F}$, $\boldsymbol{\omega}$ and $\boldsymbol{v}_\infty$ during vortex shedding past a two-dimensional bluff-body.

## 8. Conclusion

The drag of two-dimensional bodies appears to be correlated with the ratio of the kinetic energy to the vorticity in its wake, at least in the range $9 \times 10^3 \leq Re \leq 1.44 \times 10^5$. (Note, however, that with only a single data point corresponding to a turbulent wake, more empirical evidence is needed to confidently extend the correlation into the turbulent wake regime.) The fluid boundary, defined herein as any fluid possessing vorticity, specifies the volume over which the above ratio is to be evaluated. To make calculation of the ratio possible from whole-field data from the literature, all vorticity is assumed to reside in the mean wake bubble. At lower Reynolds numbers (not studied here), this assumption becomes more difficult to justify. As a limiting case of the drag correlation, ideal flows possess infinitely thin boundaries and therefore infinite vorticity in these boundaries, and hence no drag. To evaluate the correlation parameter, $G$, the mean wake bubble was modeled as a pair of elliptical patches of uniform vorticity. To aid in computing the required correlation parameter from existing test data, a Strouhal number, formulated using the elliptical vortex patch, can be used in lieu of the time-averaged velocity at a point on the vortex perimeter. The correlation indicates that the drag force on a body is proportional to the flow speed and the mass flow rate in the boundary of the fluid, i.e., the wake bubble. Absent any quantization of vortex size, this mass flow rate, and hence the drag force, can become unbounded as the scale of the contained vortices becomes finer.

## Appendix

In cases where the quantities necessary for calculating $G$ in the mean recirculation region are reported directly from experimental studies, the following discussion is unnecessary. However, since these values are rarely available, the quantities must be deduced from available measurements in the wake.

In order to relate the time-mean wake vortex to the instantaneous vortex shedding, it is important to realize that the period, $T_v$, of the pair of elliptical vortices which constitute the time-mean recirculation region is related to the period, $T_s$, of the vortex shedding. In nominally planar flow, the period, $T_v$, of the time-mean elliptical vortex can be determined by integrating (9) around any elliptical streamline (since all of them have the same period), as follows:

$$\int_0^{2\pi} d\overline{x}_2 = \int_0^{T_v} \frac{v_O}{a} dt \tag{28}$$

$$\therefore T_v = \frac{2\pi a}{v_O}. \tag{29}$$

With $n$ vortices shed downstream per shedding cycle, we also have:

$$T_v = \frac{n}{f_s} \tag{30}$$

where $f_s$ is the vortex shedding frequency. Eliminating $T_v$ from (29) and (30) suggests the following potentially useful grouping:

$$S \equiv \frac{f_s\, a}{v_O} = \frac{n}{2\pi}. \tag{31}$$

For Reynolds numbers at which vortex shedding occurs, the frequency to be used in (31) is the vortex shedding frequency; below this Reynolds number range, it is the reciprocal of the period of the standing eddies.

Solving (31) for $v_O$ gives:

$$v_O = \frac{2\pi}{n} a f_s = \frac{2\pi a}{nD} St\, v_\infty$$

$$\therefore \frac{v_O}{v_\infty} = \frac{2\pi a}{nD} St, \tag{32}$$

where $St$ is the classical definition of the Strouhal number. This relation allows the velocity ratio appearing twice in (20) to be written in terms of the shedding frequency and the semi-major axis of the time-mean elliptical wake vortex, unless a splitter plate trails the body, in which case the primary vortex shedding is eliminated.

Defining the Strouhal number in terms of the mean wake bubble allows a natural extension down to low Reynolds number flows exhibiting the pair of standing eddies. For Reynolds numbers at which vortex shedding occurs, the time-mean velocity field in the near wake still appears as a pair of counter-rotating, elliptical vortices (see, e.g., Van Dyke 1982). Below this Reynolds number range, the reciprocal of the period of the standing eddies can be used as the Strouhal frequency. As mentioned earlier, the quantity $a$ is the maximum radius of the time-mean elliptical wake vortex, and $v_O$ is the maximum velocity on the perimeter (at the minimum radius). The Strouhal number defined by (31) is equal to the characteristic vortex frequency divided by the vortex eigenvalue, $\lambda = \pm i\, v_O/a$, of the system $\dot{\boldsymbol{x}} = \boldsymbol{A}\boldsymbol{x}$ represented by (10). The near constancy of many Strouhal number groupings is due to the fact that they use length and velocity scales that are close to those of the eigenvalue of the mean wake vortex. The above relation allows the velocity field in the wake vortex to be estimated by measurements of the Strouhal number.